\documentclass[prd,twocolumn,superscriptaddress,floatfix,amsmath,amssymb,amsfonts,nofootinbib,longbibliography]{revtex4-2}

\usepackage{placeins}
\usepackage{tcolorbox}
\usepackage{float} 
\usepackage{scalerel}
\usepackage{dblfloatfix}
\usepackage[normalem]{ulem}
\usepackage[english]{babel}
\usepackage{graphicx}
\usepackage{dcolumn}
\usepackage{bm}
\usepackage{blindtext}
\usepackage{verbatim}
\usepackage{relsize}
\usepackage{mathrsfs}
\usepackage{musicography}
\usepackage{amsmath}
\usepackage{blindtext}
\usepackage{cancel}
\usepackage{physics}
\usepackage{epstopdf}
\usepackage{mathtools}
\usepackage{blindtext}
\usepackage{tensor}
\usepackage{color}
\usepackage[usenames,dvipsnames]{pstricks}
\usepackage{epsfig}
\usepackage{pst-grad} 
\usepackage{pst-plot} 
\usepackage{hyperref}
\usepackage{verbatim}
\usepackage{slashed}
\usepackage{dsfont}
\usepackage[english]{babel}
\usepackage{amsmath,amssymb,amsthm}  
\usepackage{leftindex,enumerate}
\usepackage{enumitem, kantlipsum}  

\newcommand{\mf}{\mathsf}

\newcommand{\ii}{\mathrm{i}}

\newcommand{\tc}[1]{\textsc{#1}}

\DeclareMathOperator\erfi{erfi}

\definecolor{goldenrod}{rgb}{0.85, 0.65, 0.13}

\allowdisplaybreaks[1] 

\begin{document}

\title{Cosmological Expansion Induces Interference Between Communication and Entanglement Harvesting}

\author{Matheus H. Zambianco}
\email{mhzambia@uwaterloo.ca}

\affiliation{Department of Applied Mathematics, University of Waterloo, Waterloo, Ontario, N2L 3G1, Canada}
\affiliation{Institute for Quantum Computing, University of Waterloo, Waterloo, Ontario, N2L 3G1, Canada}
\affiliation{Perimeter Institute for Theoretical Physics, Waterloo, Ontario, N2L 2Y5, Canada}

\author{Adam Teixid\'{o}-Bonfill}
\email{adam.teixido-bonfill@uwaterloo.ca}

\affiliation{Department of Applied Mathematics, University of Waterloo, Waterloo, Ontario, N2L 3G1, Canada}
\affiliation{Institute for Quantum Computing, University of Waterloo, Waterloo, Ontario, N2L 3G1, Canada}
\affiliation{Perimeter Institute for Theoretical Physics, Waterloo, Ontario, N2L 2Y5, Canada}

\author{Eduardo Mart\'{i}n-Mart\'{i}nez}
\email{emartinmartinez@uwaterloo.ca}

\affiliation{Department of Applied Mathematics, University of Waterloo, Waterloo, Ontario, N2L 3G1, Canada}
\affiliation{Institute for Quantum Computing, University of Waterloo, Waterloo, Ontario, N2L 3G1, Canada}
\affiliation{Perimeter Institute for Theoretical Physics, Waterloo, Ontario, N2L 2Y5, Canada}

\begin{abstract}

We investigate the interplay between genuine entanglement harvesting and communication mediated correlations for local particle detectors in expanding cosmological spacetimes. Focusing on a conformally coupled scalar field in de Sitter spacetime, we analyze how spacetime expansion induces interference between these two sources of entanglement when the detectors are in causal contact. We compare two physically distinct detector models: detectors whose spatial profile expands with the Universe, and detectors whose proper size remains fixed despite cosmological expansion. We find that the lack of time-reversal symmetry in cosmological settings generically leads to constructive or destructive interference between communication mediated correlations and harvested field correlations, dramatically affecting the entanglement that detectors can acquire. In particular, rapid expansion can suppress entanglement entirely for expanding detectors through destructive interference, even when both communication and field correlations are individually large, whereas detectors that maintain a fixed proper size remain capable of acquiring significant entanglement. Our results show that cosmological expansion qualitatively reshapes the balance between communication and harvesting, and that the detector internal cohesion (whether it expands with the Universe or not) plays a crucial role in determining whether detectors' entanglement can survive in rapidly expanding universes.

\end{abstract}

\maketitle

\section{Introduction}

Entanglement harvesting refers to a family of relativistic quantum information protocols in which localized particle detectors (modelled in different ways, such as Unruh-DeWitt detectors~\cite{Unruh1976,DeWitt,Unruh-Wald,Schlicht,Jorma}, atomic probes, etc.) extract entanglement from the ground state of a quantum field~\cite{Reznik2003, MenicucciEntanglingPower, HarvestingBHLaura, Pozas-Kerstjens:2015, Pozas2016,KojiMasahiroPure,boris,freefall}.



Entanglement harvesting is most prominent when the two detectors are spacelike separated. In this setup, the entanglement they acquire can only originate from pre-existing correlations in the field. However, it is not uncommon in the literature to consider entanglement harvesting for detectors that are in causal contact, therefore extracting correlations from the field in regions that can be timelike or lightlike separated. However, as first discussed by \cite{ericksonNew}, the entanglement acquired by detectors in causal contact might not be entirely extracted from pre-existing correlations in the state of the field, but rather be a result of the detectors' ability to use the field to communicate. Therefore, when the detectors are in causal contact, the authors of \cite{ericksonNew} argue that the entanglement acquired by the detectors must be split into two contributions: {\it communication-mediated} entanglement and {\it genuinely harvested} entanglement.

In \cite{ericksonNew}, the authors studied the different contributions to causal-contact entanglement harvesting in Minkowski spacetime. In this work, we propose the study of these two different contributions to the entanglement of local probes in cosmological spacetimes. Entanglement in cosmological contexts and in particular entanglement harvesting protocol in cosmological backgrounds has been explored in the literature (see, e.g. \cite{EduCosmological2012,EduNico2014,MenicucciEntanglingPower, Cosmo, collectCalling}). However, only recently, the role of communication vs genuine entanglement harvesting has been analyzed, finding that the two contributions to the acquired entanglement (communication and harvesting) can destructively and constructively interfere to either hinder or amplify the amount of entanglement the detectors can acquire while in causal contact~\cite{MatheusAdamEdu2024}. 

Here, we explore in more detail the entanglement harvesting protocol in a spatially flat Friedmann-Robertson-Walker (FRW) spacetime with a scalar quantum field conformally coupled to the curvature, paying special attention to the fact that the lack of time-reversal symmetry introduces interference between harvesting and communication contributions.  We also discuss the difference between particle detectors that expand with the Universe (the length of the detectors expands with the Universe as if there were no binding forces within it to keep it from expanding) and detectors whose internal cohesion prevents the expansion of the Universe from affecting their size (for example, atoms or small tightly bound probes). The first case is relevant when one is modeling fundamental processes and the particle detector models capture some of their features. The second is relevant because those are the physical probes we have in light-matter interaction experiments. 

In this paper, we focus on the interference between communication and harvesting in cosmological scenarios for the two kinds of detectors (those that expand with the Universe and those that remain internally bound). We will show that for the most part, they behave similarly. However, when the expansion of the Universe is fast, their ability to acquire entanglement can be very distinct due to the destructive interference between communication and harvesting.

Our manuscript is organized as follows. In Sec.~\ref{sec:Entanglement Harvesting} we review the protocol of entanglement harvesting to establish notation. In Sec. \ref{sec:Communication}, we review the definition of the different sources for the entanglement acquired by the detectors. Sec. \ref{sec:massless_FRW} describes the mathematics of an entanglement harvesting setup with a massless, conformally coupled field in a spatially flat cosmological spacetime. The main results are presented in Sec. \ref{sec:Results_deSitter}, where we perform a numerical study in de Sitter spacetime. The concluding remarks appear in Sec. \ref{sec:Conclusion}.

\section{Entanglement Harvesting}\label{sec:Entanglement Harvesting}

In this section, following the work of, e.g.~\cite{Pozas-Kerstjens:2015}, we review the protocol of entanglement harvesting to establish the notation that will be used throughout the paper. In the standard version of the protocol, one considers two Unruh-DeWitt detectors~\cite{Unruh1976,DeWitt,Unruh-Wald,Schlicht,Jorma}, A and B, following timelike trajectories $\mf z_{\tc{a}}(\tau_{\tc{a}})$ and $\mf z_{\tc{b}}(\tau_{\tc{b}})$ in an $n + 1$ dimensional spacetime $\mathsf{M}$, $\tau_{\tc{a}}$ and $\tau_{\tc{b}}$ being taken as the respective proper times of the detectors. In the detectors' proper frames, their free Hamiltonians are written as
\begin{equation}
    \hat{H}_{\tc{d}, j} = \Omega_{j} \hat{\sigma}_j^{+} \hat{\sigma}_j^{-}, 
    \label{eq:H_detectors}
\end{equation}
for $j = \text{A, B}$. Here, $\Omega_{j}$ is the proper internal energy gap, whereas $\hat{\sigma}_j^{+}$ and $\hat{\sigma}_j^{-}$ are the usual ladder operators. Taking the Hilbert space of each detector to be spanned by the orthonormal states $\{\ket{g_j}, \ket{e_j}\}$ (the ground and excited states, respectively), we can write $\hat{\sigma}_j^{+} = | e_{j} \rangle \langle g_{j}|$ and $\hat{\sigma}_j^{-} = | g_{j} \rangle \langle e_{j}|$.

The quantum field $\hat{\phi}(\mf x)$ is assumed to be a real scalar field satisfying the Klein-Gordon equation
\begin{equation}
    (\nabla^{\mu}\nabla_{\mu} - m^2 - \xi R)\hat{\phi}(\mf x) = 0,
    \label{eq:KG_eq}
\end{equation}
where $\xi$ is a constant, $m$ is the mass of the field, and $R$ is the scalar curvature of the spacetime $\mathsf{M}$. Denoting by $\{u_{\bm k}(\mf x), u_{\bm k}^{*}(\mf x')\}$ an orthonormal basis of solutions for Eq.~\eqref{eq:KG_eq}, we can write
\begin{equation}
    \hat{\phi}(\mf x) = \int{\dd^{n}{\bm k} \  (u_{\bm  k}(\mf x)\hat{a}_{\bm k} + u_{\bm k}(\mf x)^{*}\hat{a}_{\bm k}^{\dagger})},
    \label{eq:real_scalar_field}
\end{equation}
where the ladder operators $\hat{a}_{\bm k}$ and $\hat{a}^{\dagger}_{\bm k}$ satisfy the bosonic canonical commutation relations:
\begin{equation}
    [\hat{a}_{\bm k}, \hat{a}_{\bm k'}^{\dagger}] = \delta^{(n)}(\bm k - \bm k').
    \label{eq:CCR}
\end{equation}
The coupling between the quantum field and the detectors is described by the following interaction Hamiltonian density
\begin{equation}
    \hat{h}_{I}(\mf x) = \lambda \big[\Lambda_{\tc{a}}(\mf x) \hat{\mu}_{\tc{a}}(\tau_\tc{a}) + \Lambda_{\tc{b}}(\mf x) \hat{\mu}_{\tc{b}}(\tau_\tc{b})\big]\hat{\phi}(\mf x),
    \label{eq:H_I_harvesting}
\end{equation}
where $\Lambda_{j}(\mf x)$ are the spacetime smearing functions and the detector's monopole moments are given by
\begin{equation}
    \hat{\mu}_{j}(\tau_j) = e^{\ii \Omega_{j} \tau_{j}} \hat{\sigma}^{+}_{j} + e^{-\ii \Omega_{j} \tau_j} \hat{\sigma}^{-}_{j}.\label{eq:defMonopole}
\end{equation}
The time evolution in the interaction picture can be implemented using the operator
\begin{equation}
    \hat{U}_{I} = {\cal T} \exp \left( -\ii\int \dd V  \hat{h}_{\tc{I}}(\mf x)\right),
    \label{eq:U_I}
\end{equation}
where $dV$ is the invariant spacetime volume element and ${\cal T}$ is the time ordering symbol. For spatially smeared detectors one can object to the expression \eqref{eq:U_I} due to the possible ambiguity in the definition of the time ordering operation. However, \cite{EduTalesBruno2021} showed that provided that we keep all our predictions to leading order in the perturbation parameter $\lambda$ and work with suitable initial detector states, such as the ground states, then the evolution of the system will not depend on the time ordering chosen.

We consider the entanglement harvesting protocol where we prepare the initial state of the joint system with the two detectors on their respective ground states and the field in any Gaussian state $\hat{\rho}_{\phi, 0}$ with zero mean (such as the vacuum $|0 \rangle \langle 0|$, for instance). Then, the initial state can be written as
\begin{equation}
    \hat{\rho}_{0} = |g_{\tc{a}} \rangle \langle g_{\tc{a}}| \otimes | g_{\tc{b}} \rangle \langle g_{\tc{b}}| \otimes \hat{\rho}_{\phi, 0}.
    \label{eq:initial_state}
\end{equation}
With this choice, the detectors' initial state \mbox{$\hat{\rho}_{\tc{ab}, 0} = |g_{\tc{a}} \rangle \langle g_{\tc{a}}| \otimes | g_{\tc{b}} \rangle \langle g_{\tc{b}}|$} is not entangled. After interacting with the field, the detectors' state becomes $\hat{\rho}_{\tc{ab}} =  \Tr_{\phi}[\hat{U}_{I} \hat{\rho}_{0} \hat{U}_{I}^{\dagger}]$. In the basis $\{\ket{g_{\tc{a}}}\otimes \ket{g_{\tc{b}}}, \ket{g_{\tc{a}}} \otimes \ket{e_{\tc{b}}}, \ket{e_{\tc{a}}} \otimes \ket{g_{\tc{b}}}, \ket{e_{\tc{a}}} \otimes \ket{e_{\tc{b}}}\}$, the representation of $\hat{\rho}_{\tc{ab}}$ reads
\begin{equation}
 \hat{\rho}_{\tc{ab}} = \begin{bmatrix}
{ 1 - {\cal L}_{\tc{aa}} - {\cal L}_{\tc{bb}}} & 0 & 0 & {\cal M}^{*} \\
0 & {\cal L}_{\tc{bb}}& {\cal L}_{\tc{ba}} & 0 \\
0 & {\cal L}_{\tc{ab}} & {\cal L}_{\tc{aa}} & 0\\
{\cal M} & 0 & 0 & 0\\ 
\end{bmatrix} 
 +  {\cal O}(\lambda ^{4}),
\label{eq:rho_AB_harvesting}
\end{equation}
where
\begin{align}
    {\cal L}_{ij} &=  \lambda^2 \int{\dd V \dd V'e^{-\ii (\Omega_{i} \tau_i-\Omega_{j}\tau_j')} \Lambda_{i}(\mf x) \Lambda_{j}(\mf x')W(\mf x, \mf x')},\label{eq:Lij_harvesting} \\
    {\cal M} &= -\lambda^2 \int{\dd V \dd V'e^ {\ii( \Omega_{\tc{a}}\tau_\tc{a}+\Omega_{\tc{b}} \tau_\tc{b}')} \Lambda_{\tc{a}}(\mf x)\Lambda_{\tc{b}}(\mf x')G_{\tc{f}}(\mf x, \mf x')}.
    \label{eq:M_harvesting}
\end{align}
In the last equation, $G_{\tc{f}}(\mf x, \mf x')$ stands for the {\it Feynman propagator}, which can be written as
\begin{equation}
\begin{aligned}
    G_{\tc{f}}(\mf x, \mf x') &= \Tr[{\cal T}\hat{\phi}(\mf x)  \hat{\phi}(\mf x') \hat{\rho}_{\phi,0}]  \\&= \Theta(t - t') W(\mf x, \mf x') + 
    \Theta(t' - t) W(\mf x', \mf x),
    \label{Feynman_propagator}
\end{aligned}
\end{equation}
where $\Theta(t)$ stands for the Heaviside step function, and $t$ denotes any time coordinate.

Now that we have the reduced density matrix $\hat{\rho}_{\tc{ab}}$ for the detectors, we can quantify the amount of entanglement acquired by them. One popular measure in the entanglement harvesting literature (see, e.g. \cite{Pozas-Kerstjens:2015, Pozas2016, Lensing, Henderson2019}) is the {\it negativity} \cite{VidalNegativity}. This is an easy to compute entanglement quantifier that is defined even for mixed states. In the case of a system with two qubits, the negativity is defined as the absolute sum of the negative eigenvalues of the partially transposed density matrix of the two qubits $\hat{\rho}^{t_{\tc{a}}}$. For the state $\hat{\rho}_{\tc{ab}}$ described by Eq.~\eqref{eq:rho_AB_harvesting} we have~\cite{Pozas-Kerstjens:2015}
\begin{equation}
   {\cal N}=  \max\{0, {\cal V}\}+ {\cal O}(\lambda ^{4}),
   \label{eq:negativity}
\end{equation}
with
\begin{equation}
    {\cal V} = \sqrt{|{\cal M}|^{2} +\left(\frac{{\cal L}_{\tc{aa}}
    - {\cal L}_{\tc{bb}}}{2}\right)^2 } - \frac{{\cal L}_{\tc{aa}} + {\cal L}_{\tc{bb}}}{2}.
    \label{eq:V_negativity_general}
\end{equation}
Therefore, for setups where ${\cal V} > 0$, the interaction with the field causes the detectors to become entangled. When the detectors are spacelike separated, non-zero negativity means that all the entanglement acquired by the detectors is due to harvesting the pre-existing correlations in the state of the field $\hat{\rho}_{\phi}$. Otherwise, if the detectors are in causal contact, they can also get entangled via communication through the field. In the following section, we review the arguments that allow one to identify the different sources of entanglement in entanglement harvesting setups.

\section{communication-assisted entanglement vs genuine entanglement harvesting}

As first pointed out by \cite{ericksonNew}, in the study of entanglement harvesting protocols it is necessary to split the entanglement acquired by the detectors into two contributions: {\it communication-assisted entanglement} and {\it genuine entanglement harvesting}. Choosing the negativity $\mathcal{N}$ as the measure of entanglement, we follow the notation established in \cite{ericksonNew}: denote the communication contribution by $\mathcal{N}^{-}$, whereas the genuine harvesting is denoted by $\mathcal{N}^{+}$. Concretely, the {\it communication-assisted negativity} $\mathcal{N}^{-}$ can be defined as the negativity $\mathcal{N}$ computed by using only the communication-related terms in $\mathcal{M}$ (refer to Eq.~\eqref{eq:V_negativity_general}), while the {\it harvested negativity} $\mathcal{N}^{+}$ is the negativity computed using only the portion of $\mathcal{M}$ that comes from extracting the correlations in the quantum field.

Thus, the two different sources of entanglement $\mathcal{N}^{\pm}$ can be constructed by taking a closer look at how $\mathcal{M}$ depends or not on the pre-existing correlations in the field. This analysis is based on considering the symmetric and anti-symmetric parts of the Wightman function, namely,
\begin{equation}
    W^{+}(\mf x , \mf x') = \frac{1}{2}\Tr(\{\hat{\phi}(\mf x), \hat{\phi}(\mf x')\}\hat{\rho}_{\phi})
    \label{eq:W_plus}
\end{equation}
and
\begin{equation}
    W^{-}(\mf x , \mf x') = \frac{1}{2}\Tr([\hat{\phi}(\mf x), \hat{\phi}(\mf x')] \hat{\rho}_{\phi}).
    \label{eq:W_minus}
\end{equation}
In this way,
\begin{equation}
    W(\mf x, \mf x') = W^{+}(\mf x, \mf x') + W^{-}(\mf x , \mf x').
    \label{eq:Wightman_real_imag}
\end{equation}
Now, because the commutator $[\hat{\phi}(\mf x), \hat{\phi}(\mf x')]$ evaluates to a multiple of the identity operator, the anti-symmetric term $W^{-}$ does not depend on the state of the field. That is, no matter whether $\hat{\rho}_{\phi}$ is a highly correlated state or not, the contribution that $W^{-}$ provides to $\mathcal{M}$ (and therefore to the negativity) is the same. Thus, one concludes that $W^{-}$ is purely related to communication. On the other hand, $W^{+}$, which is clearly state-dependent, can be associated with the genuine harvesting of pre-existing correlations on the field's state $\hat{\rho}_{\phi}$. In particular, restricting the predictions to leading order in perturbation theory ensures that the anti-commutator does not contribute to communication \cite{martin-martinez2015}.

The next step is to use the components of the Wightman function $W^{\pm}$ to define the corresponding $\mathcal{M}^{\pm}$. To do this, one needs to formally define the symmetric and anti-symmetric parts of the Feynman propagator as
\begin{equation}
    G_{\tc{f}}^{\pm}(\mf x, \mf x') = \Theta(t - t') W^{\pm}(\mf x, \mf x') + 
    \Theta(t' - t) W^{\pm}(\mf x', \mf x).
    \label{eq:G_plus_minus}
\end{equation}
Given this, we define $\mathcal{M}^{\pm}$ using the same expression for $\mathcal{M}$ in Eq.~\eqref{eq:term_M} with $G_{\tc{f}}$ replaced by the corresponding  $G_{\tc{f}}^{\pm}$. Then, the definitions of the communication-assisted negativity $\mathcal{N}^{-}$ and the harvested negativity $\mathcal{N}^{+}$ naturally arise as
\begin{equation}
    \mathcal{N}^{\pm} = \max \{\mathcal{V}^{\pm}, 0 \} + \mathcal{O}(\lambda^4), 
    \label{eq:negativity_plus_minus}
\end{equation}
where
\begin{equation}
    {\cal V}^{\pm} =  \sqrt{|{\cal M}^{\pm}|^{2} +\left(\frac{{\cal L}_{\tc{aa}}
    - {\cal L}_{\tc{bb}}}{2}\right)^2 } - \frac{{\cal L}_{\tc{aa}} + {\cal L}_{\tc{bb}}}{2}.
    \label{eq:V_pm}
\end{equation}

\label{sec:Communication}

\section{Massless field in cosmological spacetime: conformal coupling}
\label{sec:massless_FRW}

\begin{figure*}[!t]
\includegraphics[width=0.67\textwidth,trim=20 20 20 20,clip]{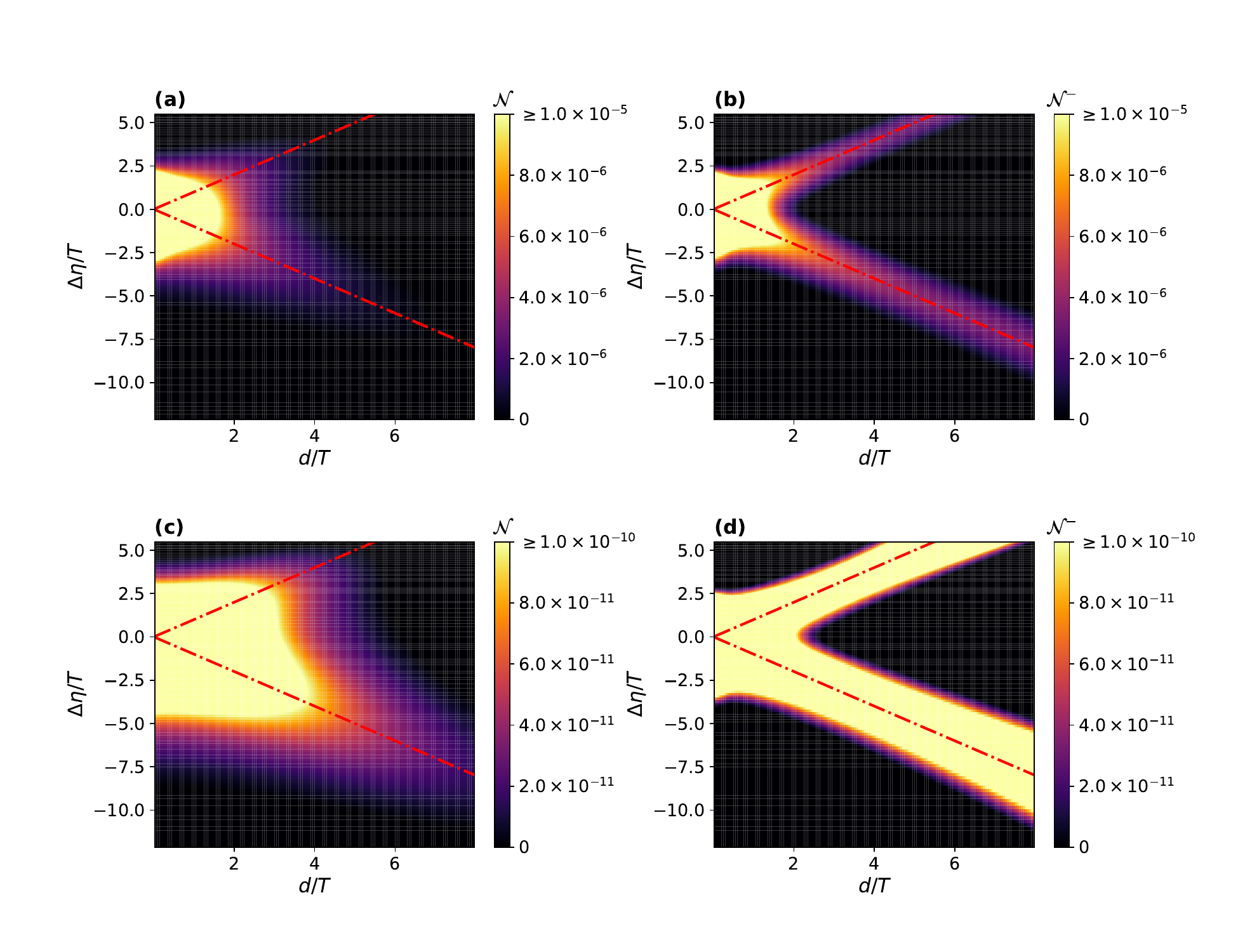}
\caption{Results for the negativity ($\mathcal{N}$) and communication-assisted negativity ($\mathcal{N}^{-}$) in de Sitter spacetime as a function of the coordinate distance $d$ between the centers of the spatial smearings of the detectors and the time delay $\Delta \eta = \eta(t_{\tc{b}}) - \eta(t_{\tc{a}} = 0) = \eta(t_{\tc{b}})$ between the centers of the switching functions. In plots (a), (b) we use $\Omega T = 4$, whereas in (c), (d) we have $\Omega T = 6$. In all cases, we set $H T = 0.1$, $\sigma/T = 0.1$. The red, dashed-dotted lines represent the light cones described by $\Delta \eta = \pm d$.}
\label{fig:grid_four_pictures}
\end{figure*}

\begin{figure*}[!b]
\includegraphics[width=0.67\textwidth,trim=20 20 20 20,clip]{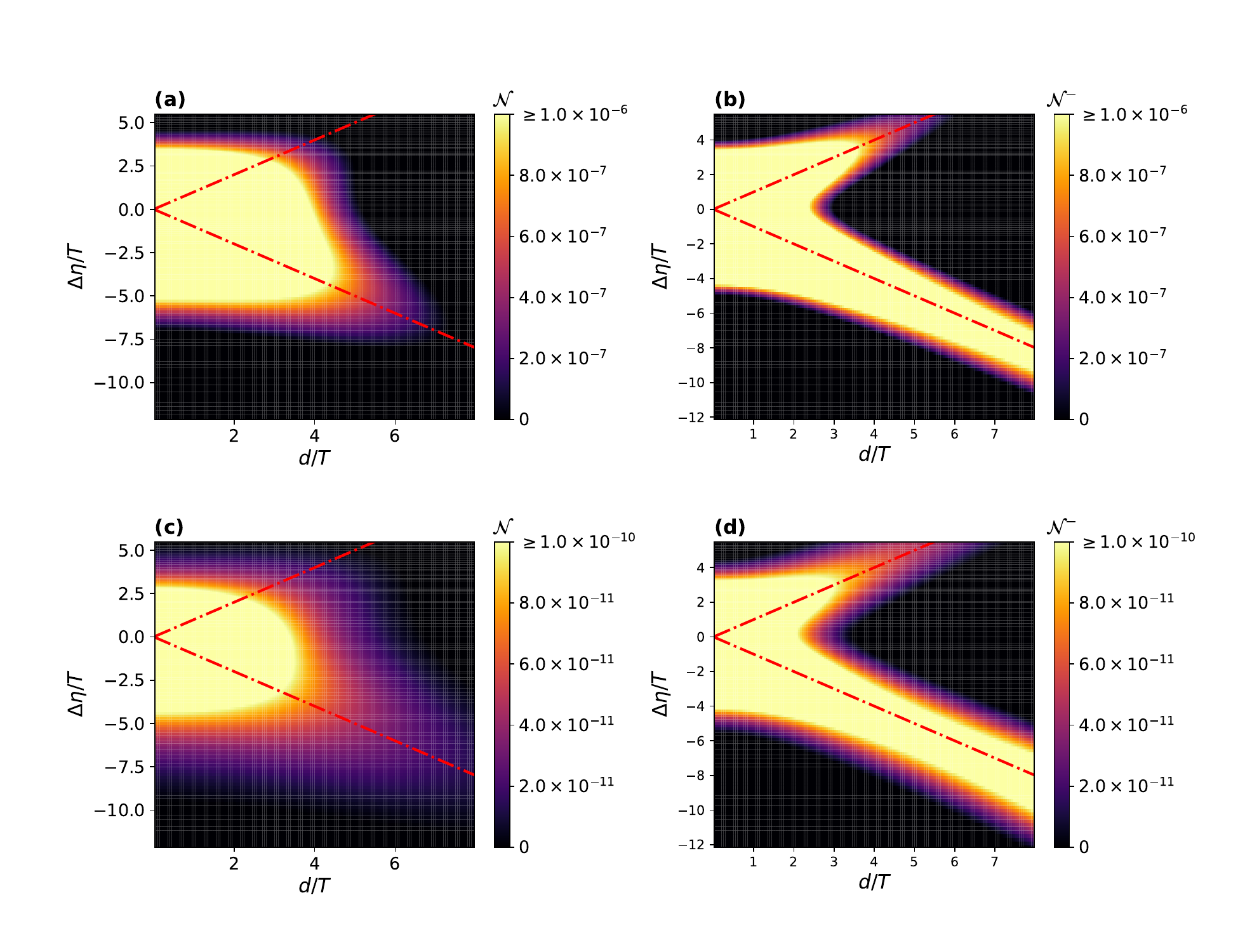}
\caption{Results for the negativity ($\mathcal{N}$) and communication-assisted negativity ($\mathcal{N}^{-}$) in de Sitter spacetime as a function of the coordinate distance $d$ between the centers of the spatial smearings of the detectors and the time delay \mbox{$\Delta \eta = \eta(t_{\tc{b}}) - \eta(t_{\tc{a}} = 0) = \eta(t_{\tc{b}})$} between the centers of the switching functions. In plots (a), (b) we use $\Omega T = 4$, and in (c)-(d) we have $\Omega T = 6$. In all cases, we set $\sigma/T = 1$ and $H T = 0.1$. The red, dashed-dotted lines represent the light cones described by $\Delta \eta = \pm d$.}
\label{fig:grid_4_pictures_2}
\end{figure*}

Adopting comoving coordinates $(t, x^{i})$, the line element of the spatially flat Friedmann-Robertson-Walker (FRW) spacetime takes the form:
\begin{equation}
    \dd s^2 = -\dd t^2 + a(t)^2\sum_{i=1}^{n}(\dd x^{i})^2.
    \label{eq:FRW_metric}
\end{equation}
Switching to conformal time, we have
\begin{equation}
    \dd s^2 = a(\eta)^2\left(-\dd \eta^2 + \sum_{i=1}^{n}(\dd x^{i})^2\right),
    \label{eq:FRW_conformal}
\end{equation}
where
\begin{equation}
    \eta(t) = \int_{0}^{t}{\dd t' \frac{1}{a(t')}}.
    \label{eq:eta}
\end{equation}
We are going to consider a massless scalar quantum field, satisfying the equation
\begin{equation}
    (\nabla_{\mu}\nabla^{\mu} - \xi R) \hat{\phi} = 0.
    \label{eq:KG_eq_1}
\end{equation}
Here, $R$ is the scalar curvature. For the sake of simplicity, here we shall consider a conformal coupling, where the scalar $\xi$ is given by
\begin{equation}
    \xi = \frac{1}{4}\frac{(n - 1)}{n}. \label{eq:xi_general}
\end{equation}
We choose to quantize the field using the conformal time:
\begin{equation}
    \hat{\phi}(\eta, \bm x) = \int{\dd^n\bm k \ (u_{\bm k}(\eta, \bm k)\hat{a}_{\bm k} + u_{\bm k}^{*}(\eta, \bm k)\hat{a}^{\dagger}_{\bm k})}. \label{eq:quantum_field_conformal}
\end{equation}
Using  $\nabla_{\mu} \nabla^{\mu} = (-g)^{-1/2} \partial_{\mu}[(-g)^{1/2}g^{\mu \nu} \partial_{\nu}]$, equation \eqref{eq:KG_eq} reads
\begin{equation}
    \left(\frac{1}{a^2}\left(\nabla^{2} - \partial^{2}_{\eta}\right) - \frac{(n - 1)\dot a}{a^3}\partial_{\eta} - \xi R \right) \hat{\phi} = 0,
\end{equation}
where the dot denotes the derivative with respect to the conformal time.

Now, for the sake of concreteness, we shall focus our attention on a $(3 + 1)$ spacetime. In this case, $\xi = 1/6$, and the scalar curvature simplifies to
\begin{equation}
    R = \frac{6}{a^3}\frac{d^2 a}{d \eta^2}. \label{eq:scalar_n=3}
\end{equation}
Then, the definition of the auxiliary field
\begin{equation}
    \hat{\Phi}(\mf x) = a(\eta) \hat{\phi}(\mf x) \label{eq:auxiliary_field}
\end{equation}
yields the equation
\begin{equation}
    (\nabla^2 - \partial^2_{\eta}) \hat{\Phi} = 0. \label{eq:KG_conformal_simplified}
\end{equation}
Therefore, this case behaves similarly to a scalar field in Minkowski spacetime, as long as the quantum field is rescaled by the expansion factor.

Now, we have everything that is necessary to compute the Wightman function. Indeed, due to relation \eqref{eq:auxiliary_field} we know that the field modes of $\hat{\phi}$ are given by
\begin{equation}
    u_{\bm k}(\eta, \bm x) = \frac{e^{\ii (\bm k \cdot \bm x - |\bm k| \eta)}  }{a(\eta)(2\pi)^{3/2}\sqrt{2|\bm k|}}.
    \label{eq:field_modes}
\end{equation}
Thus,
\begin{align}
    W(\mf x, \mf x')  & =  \frac{1}{a(\eta) a(\eta') (2\pi)^3}\int{ \frac{\dd ^{3}\bm k }{2|\bm k|} e^{\ii \bm k \cdot(\bm x - \bm x')}e^{-\ii |\bm k|(\eta - \eta')}}\nonumber\\
    & = \frac{1}{a(\eta) a(\eta') (2\pi)^2 |\Delta \bm x|}\int_{0}^{\infty}{\dd r  \sin(r|\Delta \bm x|) e^{-\ii r\Delta \eta}}
    \label{eq:W_3dim_conformal},
\end{align}
where $\Delta \eta = \eta - \eta'$ and $\Delta \bm x = \bm x - \bm x'$.  

Next, we need to obtain expressions for the terms ${\cal L}_{\tc{aa}}$, ${\cal L}_{\tc{bb}}$ and ${\cal M}$, as defined by equations \eqref{eq:Lij_harvesting} and \eqref{eq:M_harvesting}. In order to do this, we are going to choose our spacetime smearing functions to be $\Lambda_{j}(\mf x) = \chi_{j}(t) F_{j}(t, \bm x)$, with
\begin{equation}
    \chi_{j}(t) = e^{-\left( \frac{t - t_{j}}{T_{j}}\right)^2}, \quad j=A, B,
    \label{eq:gaussian_switching}
\end{equation}
and
\begin{equation}
    F_{j}(t, \bm x) = \frac{1}{a(t)^n}\frac{1}{(\sqrt{2\pi}\sigma_{j}(t))^n}e^{-\frac{1}{2}\left( \frac{|\bm x - \bm x_{j}|}{\sigma_{j}(t)}\right)^2}.
    \label{eq:gaussian_spatial_smearing}
\end{equation}
Here, $\sigma_{j}(t)$ represents the time-dependent width of the detector $j$ in comoving coordinates. Moreover, we are assuming that in the comoving coordinates the trajectories of both detectors are given by $\bm x = \bm x_{j}$, which is a constant. Moreover, notice that the definition of the spatial smearings $F_{j}$ ensures that for each time slice with $t$ constant we have
\begin{equation}
    \int{\dd^n \bm x \ a(t)^n  F_{j}(t,\bm x)} = 1. \label{eq:normalization_Fj}
\end{equation}

Now, in order to obtain analytical expressions for the terms  ${\cal L}_{\tc{aa}}$, ${\cal L}_{\tc{bb}}$ and ${\cal M}$, we can use Eq.~\eqref{eq:W_3dim_conformal} and then compute the integrals over the spatial coordinates $\bm x$ and $\bm x'$. For $i,j \in \{\text{A}, \text{B}\}$, the resulting expression can be written as
\begin{equation}
    {\cal L}_{ij} = \frac{\lambda^{2}}{2(2\pi)^3}\int{\frac{\dd t \dd t'}{a(t) a(t')} e^{-\ii \Omega(t - t') }\chi_{i}(t) \chi_{j}(t') K_{ij}(t, t')},
    \label{eq:L_ij_general}
\end{equation}
where we have defined
\begin{align}
    K_{ij}(t, t') & \equiv \int{\frac{\dd \bm k^{3}}{|\bm k|}e^{\ii \bm k \cdot (\bm x_{i} - \bm x_{j})} e^{-|\bm k|^2\Sigma_{ij}(t, t^{\prime})^{2}/2} e^{-\ii |\bm k|\Delta \eta(t, t^{\prime})}},
    \label{eq:Kij_general}
\end{align}
with $\Delta \eta(t, t') = \eta(t)- \eta(t')$ and
\begin{equation}
    \Sigma_{ij}(t, t')  = \sqrt{\sigma_{i}(t)^2 + \sigma_{j}(t')^2}.
    \label{eq:Sigma_ij}
\end{equation}
Using spherical coordinates, we can evaluate $K_{ij}$ by separately considering the cases $i \neq j$ and $i = j$. For $i \neq j$, we have
\begin{align}
    K_{ij}(t, t') & = \frac{2 \pi}{\Sigma_{ij}(t, t')d}\Biggl[I\left(\frac{\Delta \eta(t, t') + d}{\Sigma_{ij}(t, t')}\right) \nonumber \\ & - I\left(\frac{\Delta \eta(t, t') - d}{\Sigma_{ij}(t, t')}\right) \Biggr], \label{eq:Kij}
\end{align}
where $d = |\bm x_{\tc{a}} - \bm x_{\tc{b}}|$, and
\begin{align}
    I(a) =& \ii\sqrt{\frac{\pi}{2}}e^{-\frac{a^2}{2}} + \sqrt{2} {\cal D}\left(\frac{a}{\sqrt{2}} \right), \label{eq:auxiliary_Iij}
\end{align}
with ${\cal D}(x) = \frac{\sqrt{\pi}}{2} e^{-x^2}\erfi{x}$ the Dawson function, $\erfi(z) = -\ii \erf(\ii z)$ and $\erf$ the error function.

For $i = j$, we take the limit $d \to 0$ in expression \eqref{eq:Kij}, yielding
\begin{equation}
    K_{jj}(t, t^{\prime})=\frac{4 \pi}{\Sigma_{j j}\left(t, t^{\prime}\right)^2}\left[1-\frac{\Delta \eta(t,t')}{\Sigma_{j j}\left(t, t^{\prime}\right)} I\!\left(\frac{\Delta \eta(t, t')}{\Sigma_{jj}(t, t')}\right)\right].
    \label{eq:K_jj}
\end{equation}

As for the term ${\cal M}$, using Eq.~\eqref{eq:M_harvesting} we have
\begin{align}
    {\cal M} = &-\frac{\lambda^2}{2 (2\pi)^3}  \int\frac{\dd t \dd t' e^{\ii \Omega(t + t')}}{a(t)a(t')}\chi_{\tc{a}}(t) \chi_{\tc{b}}(t') \nonumber\\
    & \qquad\times\left(\Theta(t - t')K_{\tc{ab}}(t, t')  + \Theta(t' - t)K_{\tc{ab}}^{*}(t, t') \right).
    \label{eq:M_conformal_form1}
\end{align}
Implementing the Heaviside function as a nested integral and noticing that $K_{\tc{ab}}^{*}(t, t') = K_{\tc{ba}}(t', t)$ this expression is equivalent to
\begin{equation}
    {\cal M} = -\frac{\lambda^2}{2 (2\pi)^3} \int_{-\infty}^{\infty}{\dd t\int_{-\infty}^{t} \dd t' \frac{e^{\ii \Omega(t + t')}}{a(t) a(t')}\mathcal{W}_{\tc{ab}}(t, t')},
    \label{eq:term_M}
\end{equation}
where
\begin{equation}
    \mathcal{W}_{\tc{ab}}(t, t') = \chi_{\tc{a}}(t) \chi_{\tc{b}}(t') K_{\tc{ab}}(t, t') + \chi_{\tc{a}}(t') \chi_{\tc{b}}(t)K_{\tc{ba}}(t, t').
    \label{eq:W_ab}
\end{equation}

We now have all the necessary components to compute the negativity \({\cal N}\) as defined in Eq.~\eqref{eq:negativity}. Next, let us focus on deriving an expression for \({\cal M}^{-}\), which will allow us to compute the communication-assisted negativity \({\cal N}^{-}\) as well.

To compute $\mathcal{M}^{-}$, we first evaluate the anti-symmetric part of the Wightman function defined in Eq.~\eqref{eq:W_minus} (see, for example, \cite{ericksonNew} for a derivation in Minkowski spacetime):
\begin{align}
       W^{-}(\mf x, \mf x')  = &  \frac{1}{2}\langle [\hat{\phi}(\mf x), \hat{\phi}(\mf x')] \rangle \nonumber \\  
       = & \frac{i}{8\pi} \frac{\delta(\Delta \eta + |\Delta \bm x|) - \delta(\Delta \eta - |\Delta \bm x|)}{a(\eta) a(\eta')|\Delta \bm x|}. 
\end{align}
Using this expression, we can evaluate the propagator \( G_{\tc{f}}^{-} \) from Eq.~\eqref{eq:G_plus_minus}, which allows us to derive \({\cal M}^{-}\) by replacing \( G_{\tc{f}} \) with \( G_{\tc{f}}^{-} \). Alternatively, \({\cal M}^{-}\) can be obtained directly from \({\cal M}\) by substituting the function \( K_{ij} \) with its imaginary part multiplied by \(\ii\). Since \(\erfi(x)\) is real for \(x \in \mathbb{R}\), a straightforward calculation gives (for \(i \neq j\)):
\begin{align}
   K_{ij}^{-}(t, t') & = 
   \frac{2 \pi \ii}{\Sigma_{ij}(t, t') d}\Biggl[h\!\left(\frac{\Delta \eta(t, t') + d}{\Sigma_{ij}(t, t')}\right) \nonumber \\ &- h\!\left(\frac{\Delta \eta(t, t') - d}{\Sigma_{ij}(t, t')}\right) \Biggr],
\end{align}
where
\begin{equation}
    h(z) = \sqrt{\frac{\pi}{2}} \exp \left[-\frac{z^2}{2} \right].
\end{equation}
Then, we have
\begin{equation}
     {\cal M}^{-} = -\frac{\lambda^2}{2(2 \pi)^3} \int_{-\infty}^{\infty}{\dd t\int_{-\infty}^{t} \dd t' \frac{e^{\ii \Omega(t + t')}}{a(t) a(t')} \mathcal{W}^{-}_{\tc{ab}}(t, t')}, \label{eq:M_minus_FRW}
\end{equation}
with
\begin{equation}
    \mathcal{W}^{-}_{\tc{ab}}(t, t') = \chi_{\tc{a}}(t) \chi_{\tc{b}}(t') K^{-}_{\tc{ab}}(t, t') + \chi_{\tc{a}}(t') \chi_{\tc{b}}(t)K^{-}_{\tc{ba}}(t, t'). \label{eq:mathcal_W_minus}
\end{equation}

\begin{figure}[!b]
    \centering
\includegraphics[width=0.45\textwidth, trim=10 10 10 10,clip]{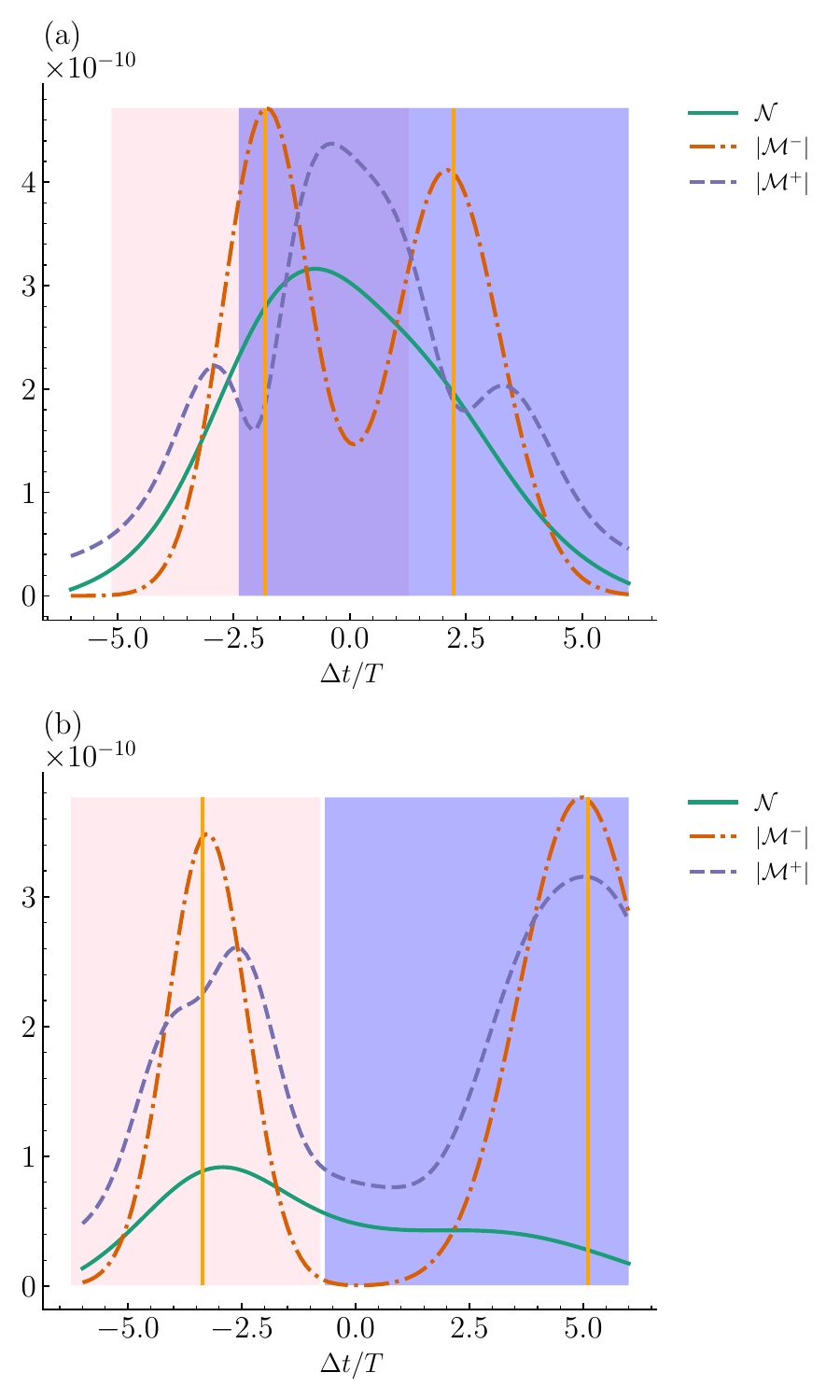}
    \caption{Negativity $\mathcal{N}$ and its different sources $|\mathcal{M}^{\pm}|$ as a function of the coordinate time delay $\Delta t = t_{\tc{b}} - t_{\tc{a}} = t_{\tc{b}}$ between the detectors, for a fixed energy gap $\Omega T = 6$ and different comoving distances (a) $d/T = 2$ and (b) $d/T = 4$. The solid, vertical yellow lines represent the light cones of detector A emanating from the event $(0, \bm x_{\tc{a}})$. The pink vertical rectangles on the left denote the region of approximate lightlike communication, with detector B in the past of detector A. Analogously, the blue vertical rectangle on the right denotes the region of approximate lightlike communication, with detector B in the future of detector A. The size of the detectors is taken as $\sigma/T = 0.1$, and the Hubble parameter is $H T = 0.1$.}
    \label{fig:1D_plots_four_cases}
\end{figure}

\begin{figure}[!b]
    \centering
    \includegraphics[width=0.45\textwidth, trim=10 10 10 10,clip]{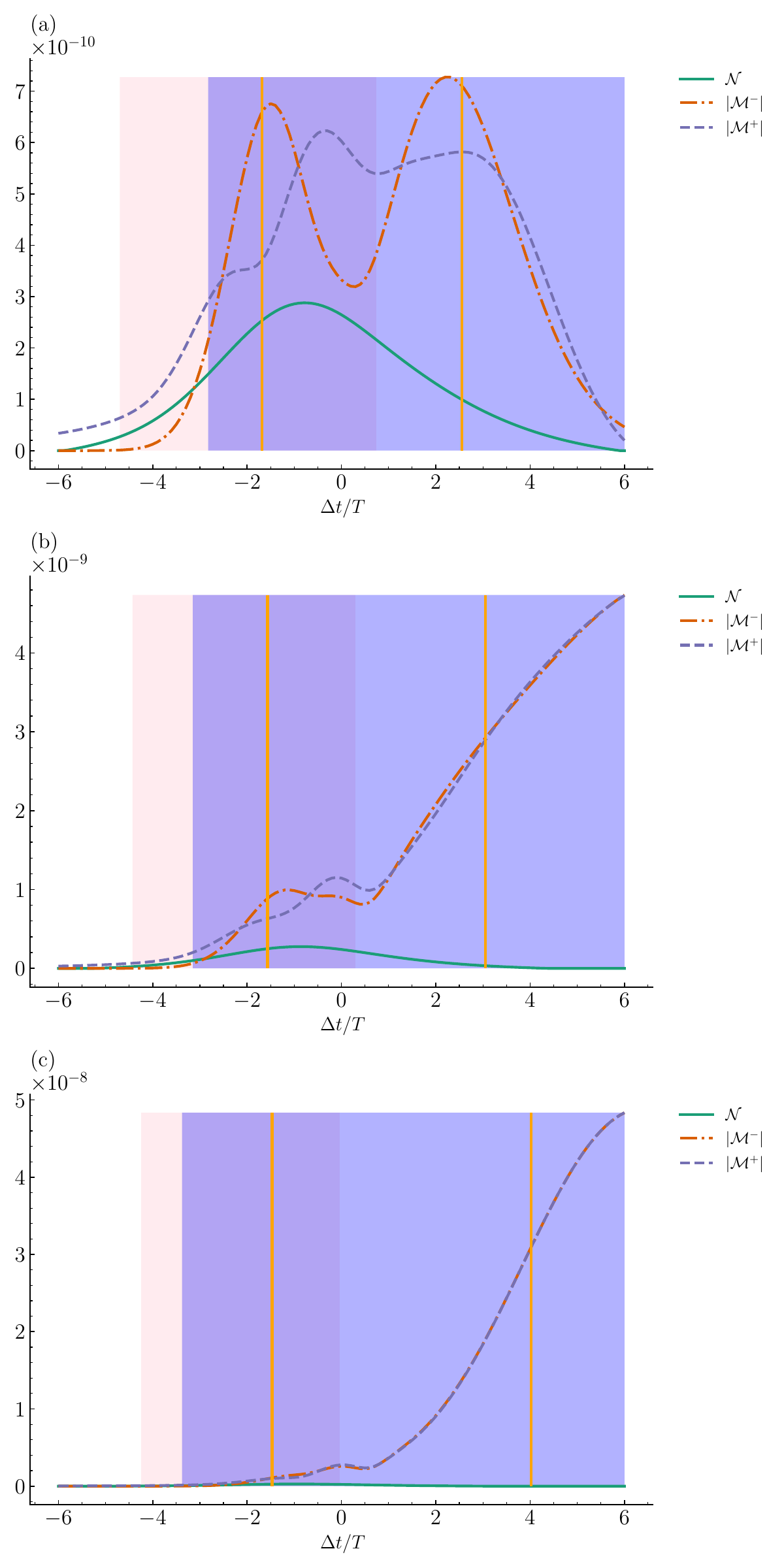}
    \caption{Negativity $\mathcal N$ and its different sources $|\mathcal{M}^{\pm}|$ as a function of the coordinate time delay $\Delta t = t_{\tc{b}} - t_{\tc{a}} = t_{\tc{b}}$ between the detectors, for a fixed energy gap $\Omega T = 6$ and comoving distance $d/T = 2$, and different Hubble parameters (a) $H T = 0.2$, (b) $ H T = 0.3$, and (c) $H T = 0.4$. The solid, vertical yellow lines represent the light cones of detector A emanating from the event $(0, \bm x_{\tc{a}})$. The pink vertical rectangles on the left denote the region of approximate lightlike communication, with detector B in the past of detector A. Analogously, the blue vertical rectangle on the right denotes the region of approximate lightlike communication, with detector B in the future of detector A. The size of the detectors is taken as $\sigma/T = 0.1$.}
    \label{fig:1D_aggressive_expansion}
\end{figure}

\begin{figure*}[!t]
    \centering
    \includegraphics[width=0.95\textwidth, trim=0 10 10 10,clip]{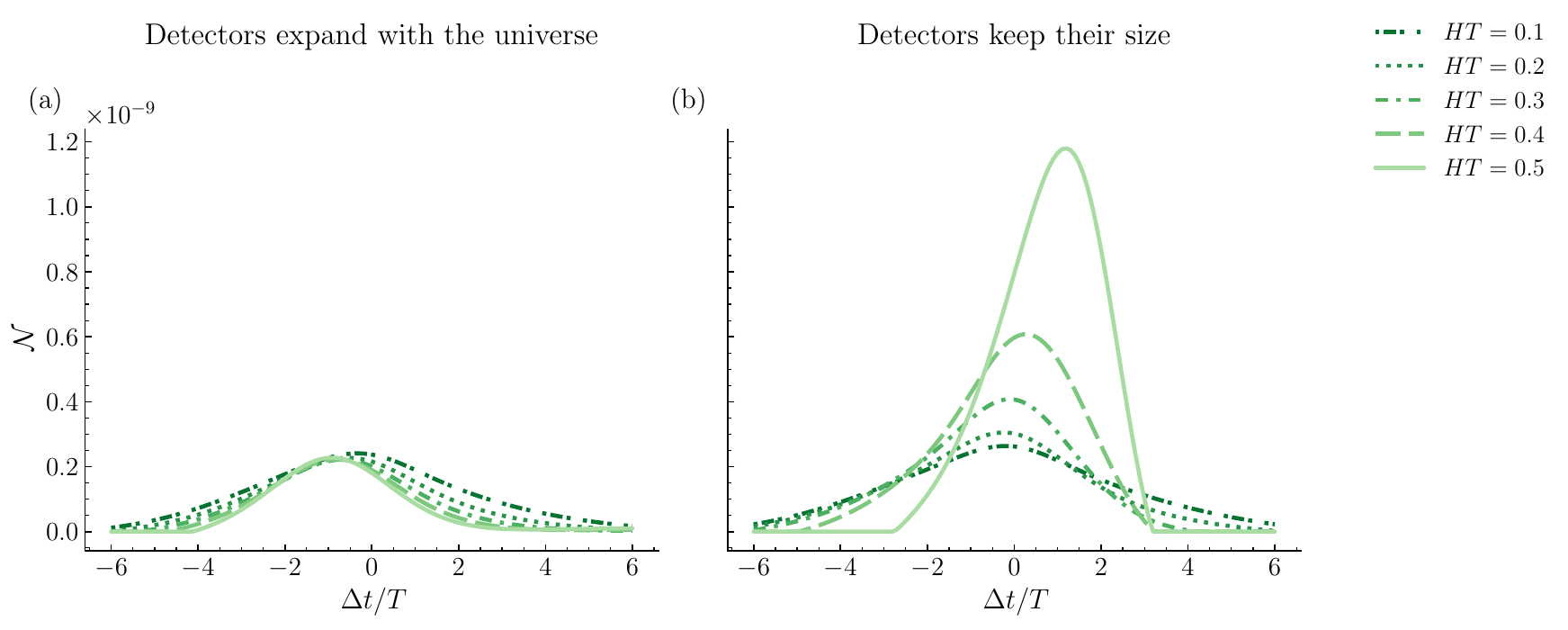}
    \caption{Negativity ($\mathcal{N}$) as a function of the coordinate time delay $\Delta t = t_{\tc{b}} - t_{\tc{a}} = t_{\tc{b}}$ between the detectors, for a fixed energy gap $\Omega T = 6$ and comoving distance $d/T = 2$, and different Hubble parameters $H T  \in \{0.1, 0.2, 0.3, 0.4, 0.5\}$. Each column represents a different detector model: in the first column (plot a) we have detectors that expand with the Universe, i.e, detectors with a constant $\sigma$. In the second column (plot b) we consider detectors that keep their size by scaling their effective size according to $\sigma(t) = \sigma/a(t)$. In both scenarios, we fix $\sigma/T=1$.}
    \label{fig:comparison_negativity_only}
\end{figure*}

\begin{figure*}[!b]
    \centering
    \includegraphics[width=0.9\textwidth]{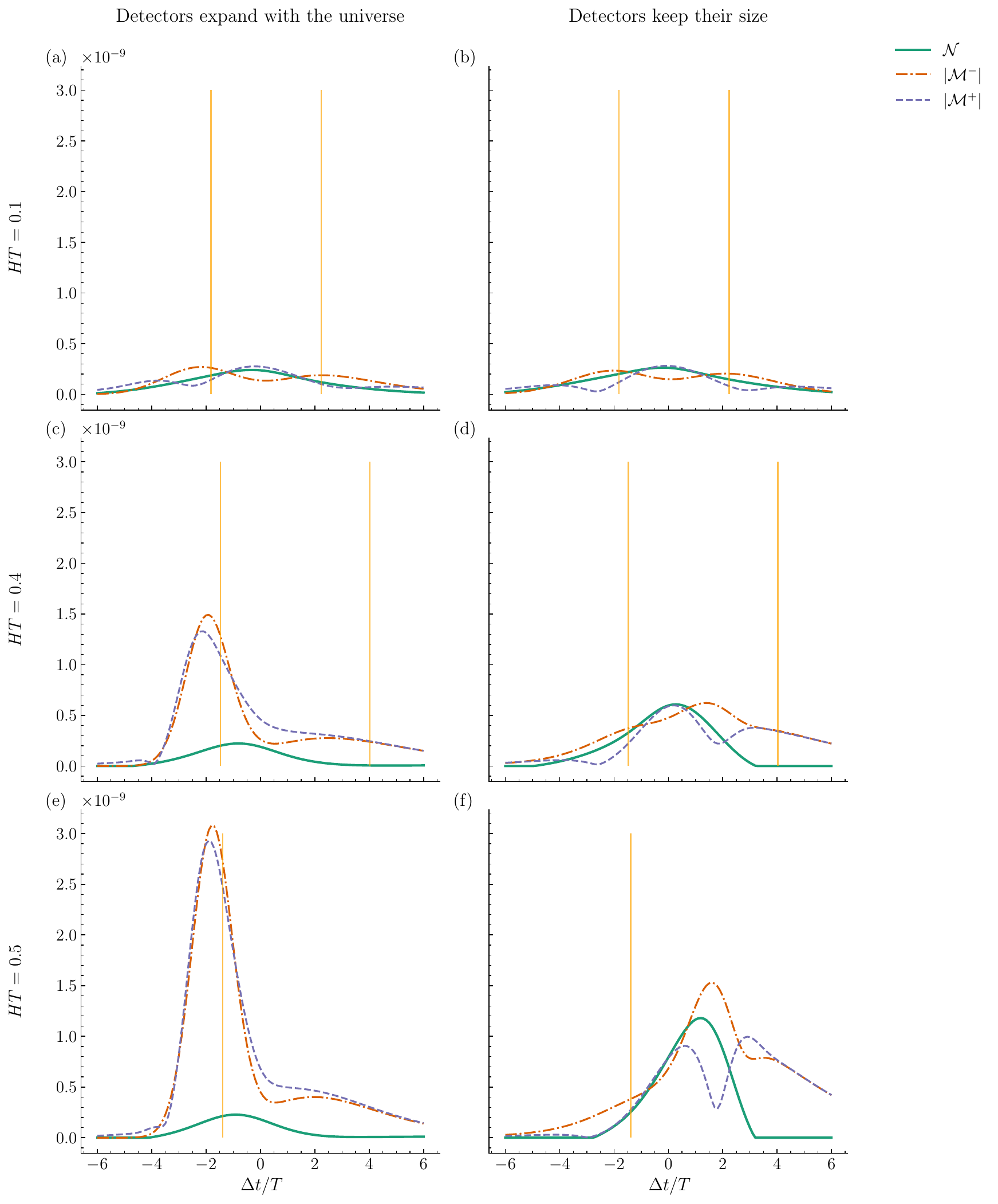}
    \caption{Negativity $\mathcal N$ and its different sources $|\mathcal{M}^{\pm}|$ as a function of the coordinate time delay $\Delta t = t_{\tc{b}} - t_{\tc{a}} = t_{\tc{b}}$ between the detectors, for a fixed energy gap $\Omega T = 6$ and comoving distance $d/T = 2$, and different Hubble parameters $H T = 0.1$, $ H T = 0.4$, and $H T = 0.5$. Each column represents a different detector model: in the first column (plots a, c, and e) we have detectors that expand with the Universe, i.e, detectors with a constant $\sigma$. In the second column (plots b, d, and f) we consider detectors that keep their size by scaling their effective size according to $\sigma(t) = \sigma/a(t)$. In both scenarios, we fix $\sigma/T=1$. The solid, vertical yellow lines represent the light cones of detector A emanating from the event $(0, \bm x_{\tc{a}})$.
    }
    \label{fig:comparison_grid}
\end{figure*}

\begin{figure*}[!b]
    \centering
    \includegraphics[width=0.95\textwidth]{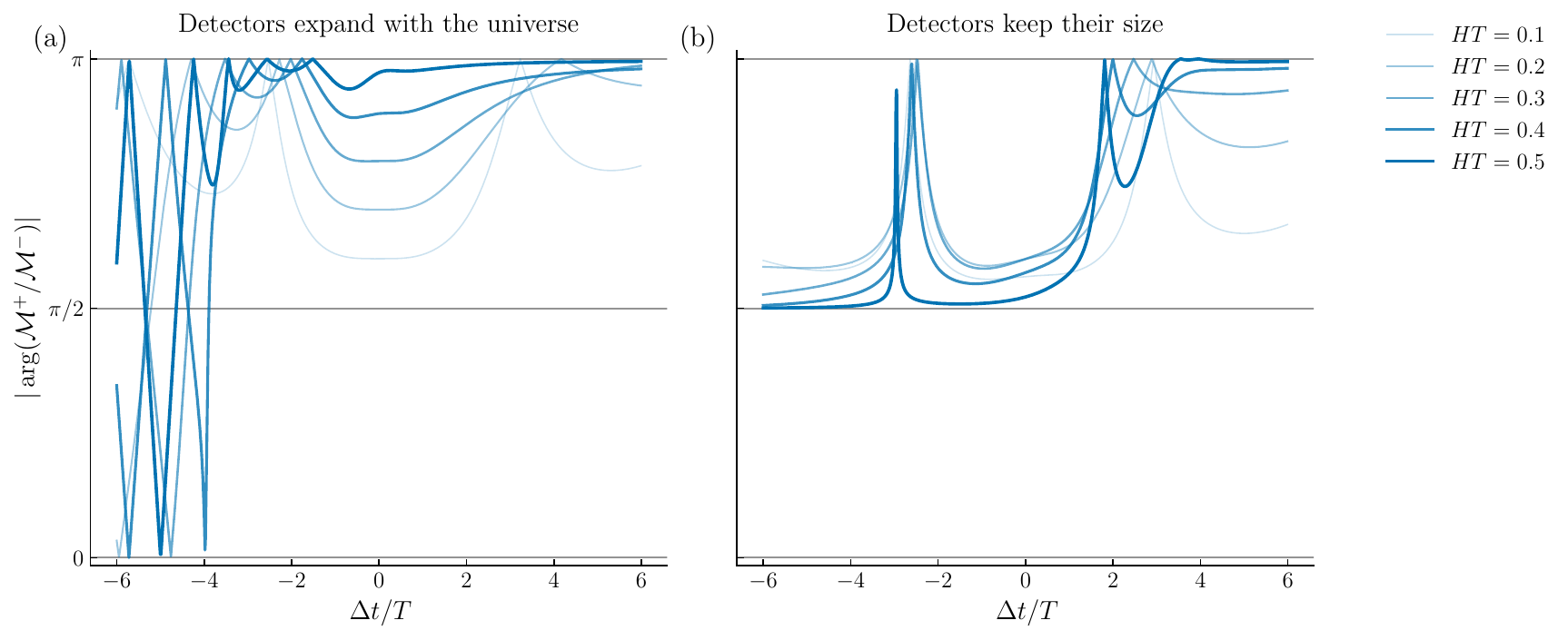}
    \caption{Absolute value of the relative phase between field correlations ($\mathcal{M}^{+}$) and communication ($\mathcal{M}^{-}$) as a function of the coordinate time delay $\Delta t = t_{\tc{b}} - t_{\tc{a}} = t_{\tc{b}}$ between the detectors, for a fixed energy gap $\Omega T = 6$ and comoving distance $d/T = 2$, and different Hubble parameters $H T  \in \{0.1, 0.2, 0.3, 0.4, 0.5\}$. Each column represents a different detector model: in the first column (plot a) we have detectors that expand with the Universe, i.e, detectors with a constant $\sigma$. In the second column (plot b) we consider detectors that keep their size by scaling their effective size according to $\sigma(t) = \sigma/a(t)$. In both scenarios, we fix $\sigma/T=1$.}
    \label{fig:comparison_grid_phases}
\end{figure*}

\section{Results in de Sitter spacetime}
\label{sec:Results_deSitter}

Now, we will use the results of section \ref{sec:massless_FRW} to explore harvesting in the particular case of de Sitter spacetime, where the expansion factor takes the form
\begin{equation}
    a(t) = e^{Ht},
    \label{eq:a_De_Sitter}
\end{equation}
and then the conformal time is given by
\begin{equation}
    \eta(t) = \frac{1}{H}\left(1 - e^{-Ht} \right).
    \label{eq:eta_De_Sitter}
\end{equation}

We will consider two cases: A) the detectors expand with spacetime, $\sigma_{\tc{a}}(t) = \sigma_{\tc{b}}(t) = \sigma$, and B) the detectors keep their size over time, which requires rescaling the comoving width of the detectors, $\sigma_{\tc{a}}(t) = \sigma_{\tc{b}}(t) = \sigma/a(t)$. Furthermore, we fix the scale of the switching functions in Eq.~\eqref{eq:gaussian_switching} to $T_{\tc{a}} = T_{\tc{b}} = T$. In what follows, we focus on expanding spacetimes, with $H > 0$.

\subsection{Detectors that expand with the Universe}
We first explore the case of detectors that expand with the Universe. Such detectors have a constant size in comoving coordinates, 
\begin{equation}
    \sigma_{\tc{a}}(t) = \sigma_{\tc{b}}(t) = \sigma.
\end{equation}
Figures \ref{fig:grid_four_pictures} and \ref{fig:grid_4_pictures_2} illustrate the results for the negativity ($\mathcal{N}$) and the communication-assisted negativity ($\mathcal{N}^{-}$). The vertical axes represent the impact of varying the time delay between the centers of the switching functions in conformal time, denoted as $\Delta \eta = \eta(t_{\tc{b}}) - \eta(t_{\tc{a}})$, with $\eta(t_{\tc{a}} = 0) = 0$, so that $\Delta \eta = \eta(t_{\tc{b}})$.
The results we show for $t_\tc{a}=0$ can be related to performing the entanglement harvesting procedure arbitrarily far in the past or future by using the results shown in Appendix \ref{apx:timeTranslation}. The Figs. \ref{fig:grid_four_pictures} and \ref{fig:grid_4_pictures_2} show the dependence of negativity on the comoving distance between the centres of the spatial smearing and the conformal time delay between the switchings for different values of the energy gap $\Omega$ and the smearing length $\sigma$. 

In Figures \ref{fig:grid_four_pictures} and \ref{fig:grid_4_pictures_2}, we observe an asymmetry with respect to the sign of $\Delta \eta$. In general, when detector B couples to the field after detector A, $\Delta \eta > 0$, the entanglement acquired by the detectors is less than in the opposite situation, $\Delta \eta < 0$. The asymmetry is  explained by the different proper distance between detectors at a fixed comoving distance $d$ being smaller for $\Delta \eta < 0$ than for $\Delta \eta > 0$, due to the expansion of the Universe.

The most important feature to be pointed out in Figures \ref{fig:grid_four_pictures} and \ref{fig:grid_4_pictures_2} is the fact that we have regions where $\mathcal{N}^{-}$ is non-zero, but the total entanglement acquired by the detectors --- as measured by $\mathcal{N}$ --- is negligible. As pointed out by \cite{MatheusAdamEdu2024}, this happens because in a general setting the communication $\mathcal{M}^{-}$ can interfere with the field correlations $\mathcal{M}^{+}$, destructively or constructively. In the situation of Figures \ref{fig:grid_four_pictures} and \ref{fig:grid_4_pictures_2}, the destructive interference causes $\mathcal{N}^{-} > \mathcal{N}$. In previous examples in the literature where the splitting between different sources of entanglement is analyzed (see, e.g. \cite{ericksonNew, Lensing, lindel2023entanglement}), such a phenomenon does not appear because of the symmetries present in the entanglement harvesting setup. Here, the lack of time reversal symmetry can be identified as the main reason behind the interference between $\mathcal{M}^{-}$ and $\mathcal{M}^{+}$ happening in de Sitter spacetime.

In Figure \ref{fig:1D_plots_four_cases}, we fix the comoving distance $d$ between the detectors to further study the relation between the $\mathcal{N}$ and $|\mathcal{M}^{\pm}|$. In Figure \ref{fig:1D_plots_four_cases}(a), around $\Delta t = 0$, communication is detrimental to the entanglement acquired by the detectors. In that region, one finds that $\mathcal {N}^+\geq \mathcal N$, which means that not including the communication contribution would increase the entanglement between detectors. Moreover, as we leave the approximate lightlike regions of communication ($|\Delta t|/T \geq 4$), the effect of communication becomes negligible and, as expected, all the entanglement acquired between the detectors is due to genuine harvesting of field correlations. In \ref{fig:1D_plots_four_cases}(b), we doubled the comoving distance between the detectors from $d/T = 2$ to $d/T= 4$. The main effect of this increase is to create a region (around $\Delta t \approx 0$) where all the entanglement acquired by the detectors is due to genuine harvesting.

In Figure \ref{fig:1D_aggressive_expansion}, we examine the effect of increasing the Hubble parameter $H$, which leads to a more rapid expansion of the Universe. We use detectors with energy gap $\Omega T =6$, effective size $\sigma/T = 0.1$, and proper separation $d/T = 2$, considering scenarios where \mbox{$H T \in \{0.2, 0.3, 0.4\}$}. Interestingly, this accelerated expansion results in the detectors not acquiring any entanglement, while the communication and pre-existing correlations contributions remain non-zero. The entanglement disappears because these two contributions interfere destructively, cancelling out almost perfectly.

\subsection{Detectors that keep their size and comparison with detectors that expand with the Universe}
We additionally consider a pair of identical detectors that keep their size (as seen from the center of mass frame of the detector~\cite{TalesBrunoEdu2020,EduTalesBruno2021}) as the Universe expands. That is, their characteristic size in comoving coordinates, denoted as $ \sigma_{\tc{a}}(t)$ and $\sigma_{\tc{b}}(t)$, obeys the relation
\begin{equation}
    \sigma_{\tc{a}}(t) = \sigma_{\tc{b}}(t) = \sigma/a(t), 
    \label{eq:sigma_time_dependent}
\end{equation}
where $\sigma$ is a constant. Here, dividing by the scale factor $a(t)$ compensates the expansion of the Universe, keeping the characteristic proper size of the detectors at $\sigma$.

Once more, we numerically evaluate the negativity $\mathcal{N}$ and its different sources, $\mathcal{M}^{\pm}$, in the concrete case of de Sitter spacetime, where the expansion factor $a(t)$ takes the form of Eq.~\eqref{eq:a_De_Sitter}. We display the results for the negativity as a function of $\Delta t/T$ in Fig.~\ref{fig:comparison_negativity_only}, where we compare the two detector models (i.e., detectors that expand with the Universe and detectors that keep their size as the Universe expands) side-by-side for increasing values of $HT$. We find that detectors that preserve their size according to Eq.~\eqref{eq:sigma_time_dependent} become more entangled than detectors that expand, even though in the latter case the amount of communication ($\mathcal{M}^-$) and field correlations ($\mathcal{M}^{+}$) considerably grows as $HT$ increases (see Fig.~\ref{fig:comparison_grid}). This can be interpreted in the context of the interference between $\mathcal{M}^{+}$ and $\mathcal{M}^{-}$ (see Ref.~\cite{MatheusAdamEdu2024}), as illustrated in Figure \ref{fig:comparison_grid_phases}, where we analyze the behavior of the phase between the components $\mathcal{M}^{+}$ and $\mathcal{M}^{-}$ as a function of $\Delta t/T$, for increasing values of $H T$. Indeed, the modulus of $\mathcal{M}$ can be written as
\begin{equation}
    |\mathcal{M}|^2 = |\mathcal{M}^{+} |^2+ |\mathcal{M}^{-}|^2 + 2 |\mathcal{M}^{-} ||\mathcal{M}^{+} | \cos \varphi,
\end{equation}
with $\varphi \equiv \arg(\mathcal{M}^{+}/\mathcal{M}^{-})$ the relative phase between $\mathcal{M}^{-}$ and $\mathcal{M}^{+}$. Because the amount of entanglement acquired by the detectors scales with $|\mathcal{M}|$ (see Eq.~\eqref{eq:V_negativity_general}), the interference between $\mathcal{M}^{+}$ and $\mathcal{M}^{-}$, as described by the variation of $\varphi$, can produce scenarios where the entanglement acquired by the detectors is significantly lower than the amount of communication and field correlations present, as depicted in Fig.~\ref{fig:comparison_grid}(e) for instance. Fig.~\ref{fig:comparison_grid_phases} seems to indicate that in the case of detectors that expand with the Universe, the usual effect of increasing $H T$ is to bring the relative phase close to $\pm\pi$, in which case the modulus of $\mathcal{M}$ assumes its lowest value possible, namely
\begin{equation}
|\mathcal{M}|_{\varphi = \pm \pi} = \sqrt{|\mathcal{M}^{+}|^2 + |\mathcal{M}^{-}|^2 - 2|\mathcal{M}^{+}||\mathcal{M}^{-}|}.
\end{equation}
On the other hand, for detectors that keep their size as the Universe expands, 
Fig.~\ref{fig:comparison_grid_phases} seems to indicate that the overall effect of increasing $H T$ is to bring the relative phase $\varphi$ closer to $\pm\pi/2$ for a wide range of values of $\Delta t/T$. In this case, we have the intermediate scenario where the sources $\mathcal{M}^{\pm}$ work together to get the detectors entangled, namely
\begin{equation}
    |\mathcal{M}|_{\varphi = \pm\pi/2}= \sqrt{|\mathcal{M}^{+}|^2 + |\mathcal{M}^{-}|^2}.
    \label{eq:interference_pi/2}
\end{equation}
Moreover, we also observe that, regardless of wether the detectors expand with the Universe or retain their shape, when $\Delta t/T \gtrsim 4$, the interference between $\mathcal{M}^{-}$ and $\mathcal{M}^{+}$ becomes destructive ($\varphi = \pm \pi$) as $HT$ increases. For $H T = 0.5$, this scenario corresponds to detector B coupling with the field when the Universe is at least around $7$ times bigger compared to when detector A coupled with the field. In contrast, for $HT = 0.1$, the relative phase tends to be closer to $\pm \pi/2$ for all values of $\Delta t/T$. This is in accordance with the fact that, for the case $H = 0$, spacetime is Minkowski, and, since we are using a time reversal symmetric switching (Eq.~\eqref{eq:gaussian_switching}), 
the phase between communication and field correlations in this case is guaranteed to be $\varphi = \pm \pi/2$ and, therefore, Eq.~\eqref{eq:interference_pi/2} holds (see Ref.~\cite{MatheusAdamEdu2024} for details).

In conclusion, detectors that keep their size as the Universe expands acquire more entanglement than detectors that expand with the Universe, even though the latter have the apparent advantage that the communication contribution to $\mathcal{M}$ is larger. This is due to the non-trivial interference between the contributions associated to communication $\mathcal M^-$ and those associated to genuine harvesting $\mathcal M^+$.

\section{Conclusions}
\label{sec:Conclusion}

We studied the splitting between communication and genuine entanglement harvesting between two detectors in causal contact in a spatially flat cosmological spacetime. The quantum field that the detectors interact with couples conformally to the curvature.  We showed that, in de Sitter spacetime, spacetime expansion introduces a non-trivial interference between the contributions to entanglement coming from communication and those associated to genuine extraction of field correlations. 

Such interference can be understood in the context of the work \cite{MatheusAdamEdu2024}, where it was shown that communication and field correlations can interfere when certain symmetries are not present in the harvesting setup. Here, the reason for this interference is the lack of time-reversal symmetry. In general, this interference can be both constructive or destructive, increasing or decreasing the entanglement acquired by the detectors in comparison with the case where no communication is allowed. 

We studied in depth two concrete interesting scenarios in de Sitter spacetime: (1) detectors that expand in tandem with the Universe, and (2) detectors keep their proper size as spacetime expands. We found that the effect of spacetime expansion is significantly different in each case. For case (1), both communication and field correlations become larger for spacetimes that expand faster. However, the faster the expansion, the more destructive the interference between those components, hindering the ability of the detectors to get entangled. In contrast, for case (2), communication and field correlations do not grow as quickly as we increase the speed of spacetime expansion. Nonetheless, the detectors become more entangled in this case, because the interference between the sources of entanglement sits more on the constructive side. Therefore, non-expanding detectors such as atoms, which would follow case (2), can get more entanglement from their interaction with a field when the Universe is exponentially expanding, due to constructive interference between harvesting and communication.

\acknowledgements

 A. T.-B. received the support of a fellowship from ``la Caixa” Foundation (ID 100010434, with fellowship code LCF/BQ/EU21/11890119). MHZ thanks Prof. Achim Kempf for funding through his Dieter Schwarz grant. Research at Perimeter Institute is supported in part by the Government of Canada through the Department of Innovation, Science and Industry Canada and by the Province of Ontario through the Ministry of Colleges and Universities. EMM acknowledges support through the Discovery Grant Program of the Natural Sciences and Engineering Research Council of Canada (NSERC). EMM thanks the support from his Ontario Early Researcher award. 

\FloatBarrier

\appendix
\section{The effect of time translation on entanglement harvesting}\label{apx:timeTranslation}

In this appendix, we study the relation between entanglement harvesting setups where the detectors couple with the field in different epochs of the Universe expansion, e.g., at the beginning of the Universe and today.

For simplicity, let us consider the case where the detectors change their size following the spacetime expansion, so that the auxiliary function of Eq.~\eqref{eq:Sigma_ij} can be taken as a constant, say $\Sigma_{ij} = 2 \sigma$. Suppose we want to perform a translation on the center of switchings as follows
\begin{align}
    t_{\tc{a}} \to  t_{\tc{a}} + \xi, \\ \nonumber
    t_{\tc{b}} \to  t_{\tc{b}} + \xi. \\
    \label{eq:translation_center_chi}
\end{align}
Then, we ask ourselves the following question: how do the physical parameters of the entanglement harvesting setup need to change in order to preserve the value of the negativity? Concretely, we would like to find expressions for
\begin{align}
    \Tilde{\sigma} & = \Tilde{\sigma}(\sigma, \xi), \\ \nonumber
    \Tilde{d}  & = \Tilde{d}(d, \xi) 
    \label{eq:tildas}
\end{align}
to ensure the conditions
\begin{equation}
    {\cal L}_{jj}(t_{j}, \sigma)  = {\cal L}_{jj}(t_{j} + \xi, \Tilde{\sigma}) ,\label{eq:Ljj_invariance_DeSitter}
\end{equation}
\begin{equation}
    |{\cal M}(t_{\tc{a}},  t_{\tc{b}}, \sigma, d)| = |{\cal M}(t_{\tc{a}} + \xi,  t_{\tc{b}} + \xi, \Tilde{\sigma}, \Tilde{d})|.  \label{eq:M_invariance_DeSitter}
\end{equation}
Let us begin with the analysis of the local noise, ${\cal L}_{jj}$. First of all, notice that if we want this term to be invariant under the proper time translation $t_{j} \to t_{j} + \xi$, it is sufficient to consider the change of variables $(t, t') \to (s, s') = (t - \xi, t' - \xi)$ on Eq.~\eqref{eq:L_ij_general}. Then, we observe that
\begin{align}
    a(t)a(t') & = e^{2H \xi}a(s)a(s'), \\
    \Delta \eta(t, t')  & = e^{-H \xi} \Delta \eta(s, s').
\end{align}
The last relation gives us a hint about how the term $K_{jj}(t, t')$ (refer to Eq.~\eqref{eq:K_jj}) should change if we want to ensure condition \eqref{eq:Ljj_invariance_DeSitter}. Indeed, by imposing that
\begin{equation}
    \frac{\Delta \eta(t, t')}{\Tilde{\sigma}} = \frac{\Delta \eta(s, s')}{\sigma}, 
\end{equation}
we get the relation
\begin{equation}
    \Tilde{\sigma} = e^{-H \xi} \sigma.
    \label{eq:sigma_tilda}
\end{equation}
Using this definition, it follows that
\begin{equation}
    K_{jj}(t, t'; \Tilde{\sigma}) = e^{2 \xi H} K_{jj}(s, s'; \sigma).
    \label{eq:K_jj_tilda_relation}
\end{equation}
Notice that the extra factor of $e^{2 \xi H}$ cancels out the one that comes from the expansion factors when written in terms of the new variables $s$ and $s'$. Therefore, by rescaling the size of the detectors via equation \eqref{eq:sigma_tilda}, we ensure that condition \eqref{eq:Ljj_invariance_DeSitter} holds.

As for the correlations term ${\cal M}$, once more we perform the change of variables $(t, t') \to (s, s') = (t - \xi, t' - \xi)$ in Eq.~\eqref{eq:term_M}. Explicitly, we can write (taking $T_{\tc{a}} = T_{\tc{b}} = 1$ to simplify the notation)
\begin{align}
   & \frac{e^{\ii \Omega(t + t')}}{a(t)a(t')} e^{-(t - t_{\tc{a}} - \xi)^2}e^{-(t' - t_{\tc{b}} - \xi)^2} \nonumber \\
   & \quad \times \left(\Theta(t - t')K_{\tc{ab}}(t , t'; \Tilde{\sigma}, \Tilde{d}) +\Theta(t' - t)K_{\tc{ba}}(t', t;  \Tilde{\sigma}, \Tilde{d})\right)  \nonumber \\
   & = e^{2 \ii \Omega \xi}e^{-2 H \xi} \frac{e^{\ii \Omega(s + s')}}{a(s)a(s')} e^{-(s - t_{\tc{a}})^2}e^{-(s' - t_{\tc{b}} )^2}\nonumber \\
   & \quad \times \left(\Theta(s - s')K_{\tc{ab}}(s + \xi , s' + \xi; \Tilde{\sigma}, \Tilde{d}) \nonumber \right.\\
   & + \left. \Theta(s' - s)K_{\tc{ba}}(s' + \xi, s + \xi;  \Tilde{\sigma}, \Tilde{d})\right).
\end{align}

Next, notice that if we make the replacement $\sigma \to \tilde{\sigma}$ as in Eq.~\eqref{eq:sigma_tilda} and define
\begin{equation}
    \Tilde{d} = e^{-\xi H} d,
    \label{eq:d_tilda}
\end{equation}
then the effect on each term $K_{\tc{ab}}$ and $K_{\tc{ba}}$ is the same as before, i.e.
\begin{equation}
    K_{ij}(t + \xi, t' + \xi; \Tilde{\sigma}, \Tilde{d}) = K_{ij}(s, s' + \xi; \sigma, d).
\end{equation}
Of course, we still have the extra phase $e^{2\ii \Omega \xi}$, but this does not affect the absolute value of the ${\cal M}$, which is the quantity that matters for computing negativity. 

Therefore, if one wants the negativity to be invariant by a joint translation of the center of the switching of the detectors, $t_{j} \to t_{j} + \xi$, then we need to rescale the parameters $\sigma$ and $d$ by the expansion factor evaluated at the translation, that is, $d \to d/a(\xi)$ and $\sigma \to \sigma/a(\xi)$.

\bibliography{references}

\end{document}